\documentclass{aa}  
\usepackage{graphicx}
\usepackage{psfig}
\usepackage{txfonts}
\begin{document}
   \titlerunning{An X-ray look at the Piccinotti AGN Mrk 590}
   \authorrunning{A.L. Longinotti et al.}
   \title{An X-ray look at the Seyfert 1 Galaxy Mrk 590: {\it XMM-Newton} and {\it Chandra} reveal complexity in circumnuclear gas}


   \author{A.L. Longinotti \inst{1}, S. Bianchi \inst{1}, M. Santos-Lleo \inst{1}, P. Rodr{\`i}guez-Pascual \inst{1}, M. Guainazzi \inst{1}, M. Cardaci
          \inst{1,2}
          \and
          A. M. T. Pollock \inst{1}
          }

   \offprints{A.L. Longinotti}

   \institute{1 European Space Astronomy Centre,  Apartado 50727 E-28080 Madrid, Spain \\
              2 Departamento de F\'{\i}sica Te\'orica, C-XI, Universidad Aut\'onoma de Madrid, 28049 Madrid, Spain \\
               \email{alonginotti@sciops.esa.int}
                       }

   \date{Received August 2006;}

 
  \abstract
   {This paper reports on a partially simultaneous observation  of the  bright Seyfert~1 Galaxy    Mrk~590,  performed  by  XMM-Newton and  Chandra.  
     }
   {The long exposure ($\sim$ 100 ks) allows to investigate with great detail the 
   Fe K complex at 6-7 keV and the presence of soft X-ray spectral features.}
   { We have analysed XMM-Newton data from  the European Photon Imaging Camera (EPIC)  in the 0.5-12~keV band and from the  Reflection Grating Spectrometer (RGS) in the 0.35-2.5~keV band, and data from  the High Energy Transmission Gratings (HETGs) onboard Chandra.   UV and optical data from the Optical Monitor (OM) onboard XMM-Newton  are also included in the analysis.}
   {The broad band spectrum is well described by an unabsorbed  power law and  three unresolved Fe~K  lines in the 6-7 keV range.
   The presence of a Compton reflection component  and a narrow Fe K~line at 6.4~keV is consistent with  an origin via  torus reflection.  
   The   ionised Fe lines  at $\sim$6.7 and 7~keV  are instead most likely  originated  by scattering on a warm and ionised gas. 
   The soft X-ray spectrum appears to be almost featureless due to the very bright continuum emission, except for  one emission  line identified as OVIII~Ly$\alpha$  detected at $\sim$19~$\AA$ by both RGS and Chandra-MEG.
The emerging picture consists of an active nucleus seen directly on a ``clean'' line of sight without intervening material, surrounded by  photoionised circumnuclear gas  at a high ionisation level. 
  We also  study three serendipitous sources in the field of view of  Chandra and  XMM-Newton.   
  One of these sources may  be identified with an ULX of L$_{0.3-10 keV}$$\sim$4$\times$10$^{40}$ ergs/s. }
   {}


   \maketitle
%

\section{Introduction}
Active Galactic Nuclei have been observed  in the X-ray domain for many years. Nowadays,   it is fair to say that  the overall spectral shape  of Seyfert~1 Galaxies is quite well-known.
The hard X-ray continuum  (2-10 keV)  is well described by  a power law produced by Inverse Compton scattering (Haardt \& Maraschi 1993) 
with average photon index  $\Gamma$~$\sim$1.9 (Piconcelli et al. 2005).
The extrapolation of the power law to  lower energies in some cases implies  
an excess in soft X-ray emission; in many sources the soft X-ray spectrum is affected  by absorption or emission features originated in a photoionised medium along the line of sight (Kaastra et al. 2000, Kaspi et al. 2001). 

Reflection of  X-ray photons by optically thick material 
may also occur and in this case it gives rise to a hard X-ray Compton reflection component and prominent Fe~K features (George \& Fabian 1991). 
The properties of Fe~K features can  provide  much information on the origin of the X-ray emitting gas.
In the case of  reflection by an  accretion disc,  the  Fe line profile is relativistically broadened and skewed  (Fabian et al. 2000).
When X-ray photons are reflected by distant material, like  the torus proposed  in the Unification Scheme (Antonucci 1993, Ghisellini, Haardt \& Matt, 1994), the resulting profile is narrow and, in most cases,  unresolved by the current instrumentation. 
Since Fe~K lines could also be produced in the Broad Line Region (BLR) in  an analogous way to  the broad UV/Optical lines, a contribution of the BLR emission to the cores of narrow Fe lines cannot be excluded (Yaqoob \& Padmanabhan 2004, Nandra 2006).

What is as yet unclear is which scenario  is the most common in Sy1 Galaxies.
Recent observations with {\it XMM-Newton} and {\it Chandra}  have shown  that a narrow Fe~K line is an almost ubiquitous component in the X-ray spectra of type-1 AGNs 
(Yaqoob \& Padmanabhan 2004, Jimenez-Bailon et al. 2005). 
On the other hand, only a handful of sources seem to host a relativistically broadened  diskline (Nandra et al. 1999, Turner et al. 2002, Ponti et al. 2002, Longinotti et al. 2003).

Mrk~590 ({\it z}=0.026)  is a bright Seyfert 1 galaxy.
It was observed by the {\it Einstein} observatory  (Kriss et al. 1980) and by  HEAO 1 as  part of the Piccinotti sample (Piccinotti et al. 1982).
{\it Exosat} data of Mrk~590 did not reveal any particular spectral complexity (Turner \& Pounds 1989).
 More recently, it was detected in the high-energy domain by {\it BATSE}, in the 20-200 keV band (Malizia et al. 1999).   
 Despite of being part of one of the best studied AGN sample (the Piccinotti's), Mrk~590 has not been observed by {\it ASCA} nor by {\it BeppoSAX}.
Therefore, the hard X-ray spectral properties of  this AGN have remained almost unknown until  the analysis  of a 10~ks  {\it XMM-Newton} observation
was reported by  Gallo et al. (2006). In this work, 
the  0.3-10~keV  flux  was reported to be about  8.4$\times$10$^{-12}$ ergs cm$^{-2}$ s$^{-1}$,  the 2-10~keV luminosity  was  0.7$\times$10$^{43}$ ergs~s$^{-1}$   and the presence of a strong Fe~K line was revealed in the EPIC data. 
Here, we will discuss  a quasi simultaneous  observation of this source   performed by {\it XMM-Newton} and {\it Chandra} for about 100~ks. 

\section{Observation and data reduction}
\subsection{The {\it XMM-Newton} data}
Mrk 590 was observed  by {\it XMM-Newton} on  July 4, 2004 (OBSID 0201020201). Data from the EPIC, RGS and OM  instruments  were available (Struder et al. 2001; Den Herder et al.  2001, Mason et al. 2001).
The nominal duration time was 100 ks. 
The observation was performed  in Small Window mode for the pn and the MOS1 cameras, while the 
MOS2 was operated in Full Frame. 
The raw data were processed with SAS 6.5.0 with the tasks \texttt{epchain} for the pn data and  \texttt{emchain} for the MOS data.
During the satellite operation, the scheduled observation was stopped for about 15 minutes 
due to telemetry loss, so that two distinct event lists were obtained for each detector.
These two event files were then merged in one file by running the SAS task \texttt{merge} 
(Gabriel et al. 2004).
Screening for flaring background due to high-energy particle was applied to this final event list. Intervals affected by background flares were removed by selecting events with a count rate lower than 0.6~cts/s  in the 10-12~keV light curve.
Source and background  counts were extracted from circular regions of 40 arcsec.
  No pile-up  affects  the observation, as tested with the task \texttt{epatplot},  so that 
pattern 0 to 4 were selected for the pn spectra and pattern 0 to 12 were selected  for the MOS ones.
The effective exposures times  are 71, 85 and 90 ks respectively for the pn, MOS1 and MOS2 (see Table  \ref{tab:log}).
Because of the different observing modes of the MOS data, it was not possible to co-add the spectra and therefore they are kept and analysed separately. 
 Spectra were grouped in order to have 100 counts/bin for the pn data and 50 counts/bin for the MOS.  
 Response matrices were created employing the SAS tasks \texttt {arfgen} and \texttt{rmfgen} for EPIC data.
 
 RGS data were processed with the task \texttt {rgsproc}, yielding an useful exposure 
 time of 105 ks.
A response matrix 
was created  for  the spectral analysis
by the task \texttt{rgsrmfgen}.
  No binning is applied to the RGS spectra.  
  The  data from the  Optical Monitor were reduced running the task \texttt{omichain}. 
\begin{table}            
\centering                     
\begin{tabular}{c c c c c}
\\      
\hline\hline                
 Satellite   & Instrument &  Exposure &  Flux$_{2-8}$ \\    
  - &   - & (ks) & (10$^{-12}$ ergs cm$^{-2}$ s$^{-1}$)    \\
\hline\hline 
\\ 
 Chandra  &  HETG & 95 & 6.72$^{+0.38}_{-0.45}$  \\
 \hline 
\\                     
XMM-Newton &     EPIC/pn & 71  & 5.62$\pm$0.09  \\
    &   EPIC/MOS1 &   85   & 5.97$\pm$0.09 \\
&   EPIC/MOS2  &   90  &  5.60$\pm$0.09 \\
\\
&   RGS  &   105    & -\\

\hline                                  
\end{tabular}
\caption{\label{tab:log}Observation Log. Mrk 590 was observed  by {\it Chandra} on July~3rd, 2004 and by {\it XMM-Newton} on the following day. A simultaneous exposure of  65 ks was obtained. }
\end{table}

\begin{figure}
   \centering
   \includegraphics[width=9cm]{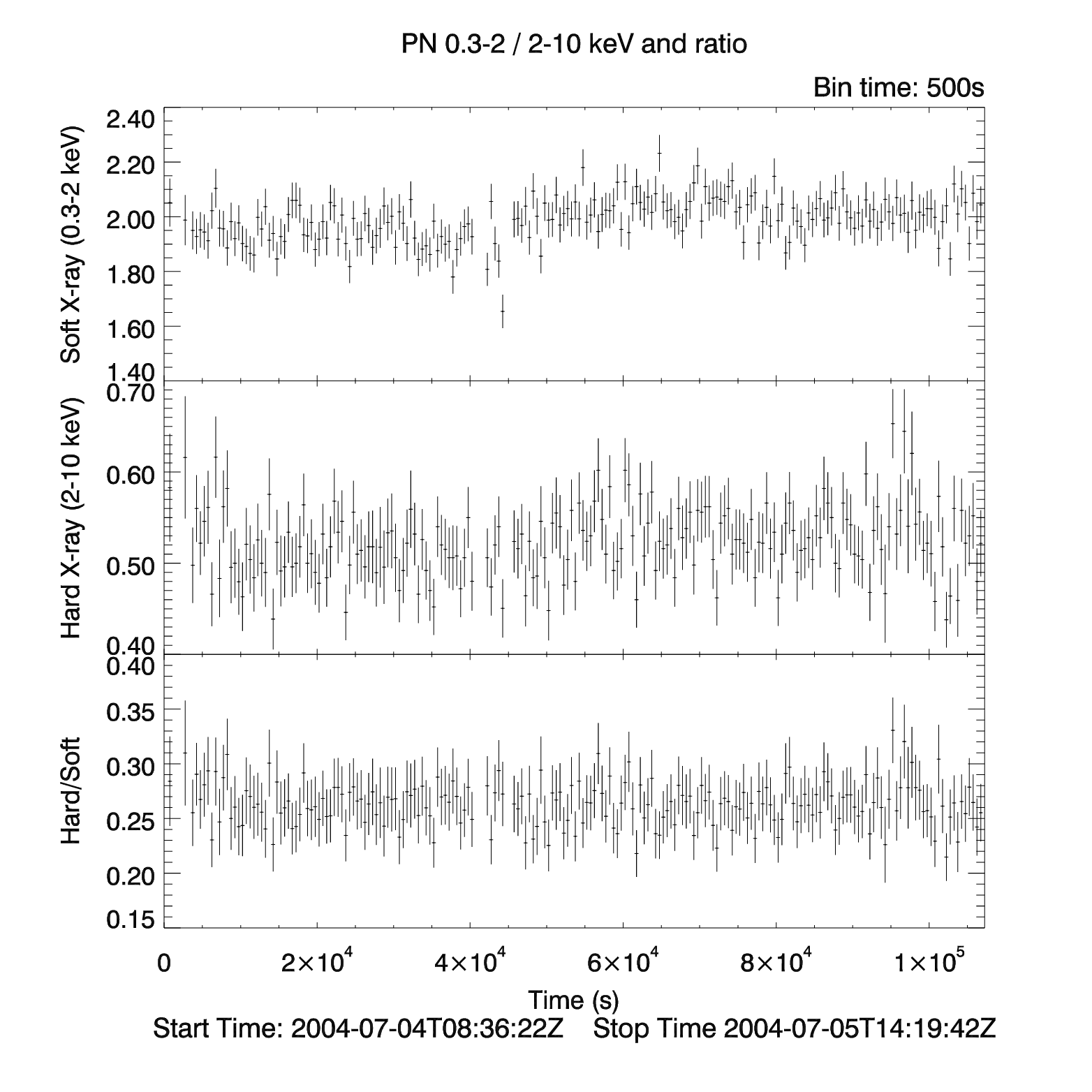}
   \caption{Light curves of the EPIC/pn data in the soft X-ray band (top), hard X-ray band (medium) and their hardness ratio (bottom). The background-subtracted  count rates are plotted.  The source does not show variability.}
              \label{fig:light}%
    \end{figure}

During the {\it XMM-Newton} observation, the source did not vary significantly in flux and it did not   show spectral variations.
The light curves of the pn data plotted in Fig. \ref{fig:light}   do not reveal  any 
particular temporal behaviour. 
Therefore, the spectral analysis described in the following sections  is performed on the integrated spectrum.

\subsection{The {\it Chandra} data}
Mrk~590 was observed by \textit{Chandra} HETG (High Energy Transmission Gratings: Canizares et al. 2005) in July 2004 for about 100 ks. The {\it Chandra HETGS} consists of two assemblies, the High Energy Grating (HEG)
in the 0.8-10~keV, and the Medium Energy Grating (MEG), in the 0.4-5~keV.
Data were reduced with the Chandra Interactive Analysis of Observations (\textsc{ciao}) 3.2.1 and the Chandra Calibration Database (\textsc{caldb}) 3.0.1, adopting standard procedures. In particular, a new evt2 file was created with \texttt{acis\_process\_events}, adopting an observation-specific bad pixel file and reconstructing the gratings extraction region and events with \texttt{tgdetect}, \texttt{tg\_create\_mask} and \texttt{tg\_resolve\_events}. 
An exam of the light curve confirms that no variability is observed during the Chandra exposure, as  already emerged for the {\it XMM-Newton} data. 
First-order HEG and MEG spectra were extracted with \texttt{tgextract} and then co-added with \texttt{add\_grating\_orders}. In the following analysis, we binned the spectra at the resolution of the instruments, i.e. 0.012 and 0.023 $\AA$ (FWHM), for HEG and MEG, respectively.
The C-statistic was employed in the spectral fitting and all statistical errors are given at 90\% level of confidence ($\Delta$C=2.71).

\section{Spectral analysis}
\subsection{{\it XMM-Newton}/EPIC spectra}
\label{subsec:epic}
Galactic absorption of column density N$_H$=2.7$\times$ 10$^{20}$cm$^{-2}$ is included in all the following spectral fits (Dickey \& Lockman, 1990). 
Errors are quoted at the 90\% confidence level for one interesting parameter (i.e. $\Delta\chi^2$=2.71).
 Throughout the text,   the energies are quoted in the source frame and the adopted cosmological constant is 
H$_0$=70~km~s$^{-1}$Mpc$^{-1}$. The measured flux and luminosity in the soft and hard X-ray band
are: F$_{0.3-2keV}$$\sim$4.4$\times$10$^{-12}$ ergs cm$^{-2}$ s$^{-1}$,  L$_{0.3-2keV}$$\sim$0.67$\times$10$^{43}$~ergs~s$^{-1}$ and F$_{2-10keV}$$\sim$6.4$\times$10$^{-12}$ ergs cm$^{-2}$~s$^{-1}$, L$_{2-10keV}$$\sim$0.97$\times$10$^{43}$~ergs~s$^{-1}$.  
\begin{figure}
   \centering
   \includegraphics[angle=-90,width=8cm]{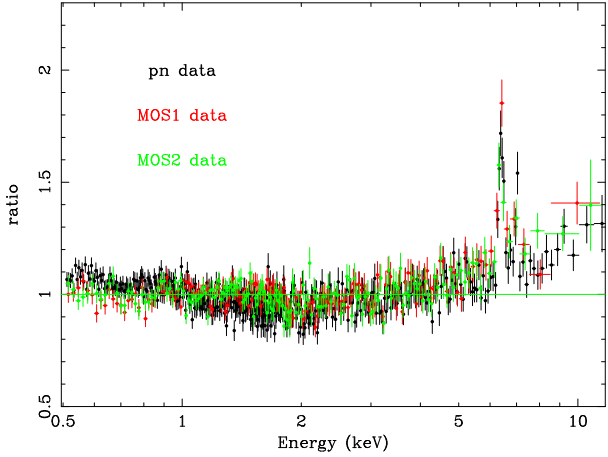}
   \caption{Ratio of the broad band  spectrum fitted by a power law with $\Gamma$$\sim$1.76. The residuals above 5 keV are evident (source frame plot). }
              \label{fig:epic_ratio}%
    \end{figure}

    \begin{figure}
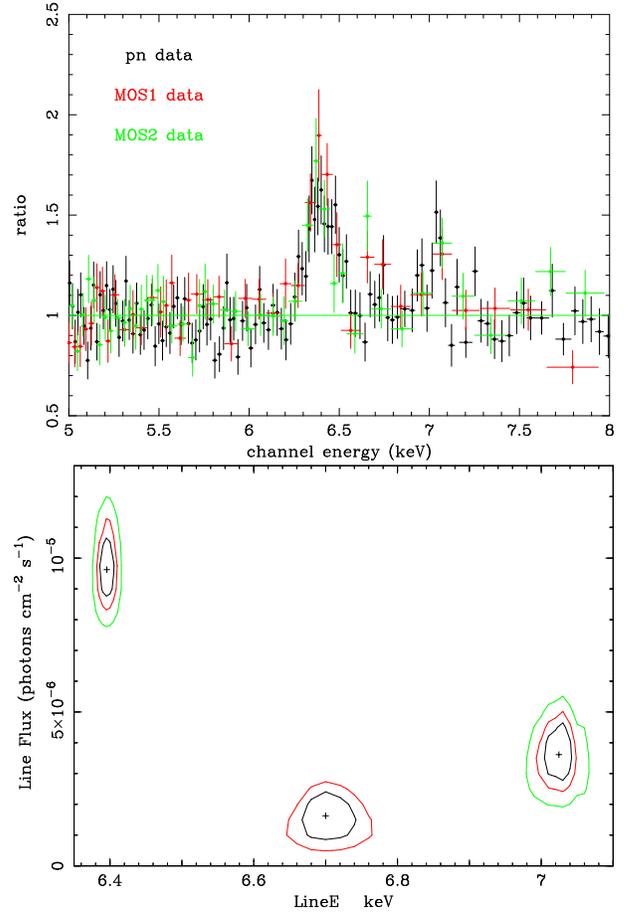

   \centering
\includegraphics[angle=-90,width=8cm]{fig3.ps}
 \includegraphics[angle=-90,width=8cm]{fig4.ps}
      \caption{ {\it Top panel}:  the ratio of the  EPIC data  fitted by a power law  shows three emission lines corresponding to Fe K transitions at $\sim$~6.4, 6.7 and 7~keV
      (source frame plot).   
     {\it Bottom panel}: Confidence contours at 68.3, 90 and 99\% levels of confidence for the energy and intensity of the 3 lines.  All the three lines have unresolved width. The line at 6.7 keV is less significant than the other two and this is why only two contours are shown (68.3 and 90\%).  }
         \label{fig:lines}
   \end{figure}
 \begin{table}     
EPIC 2-12 keV best-fit: plaw+3 zgauss \\       
\centering                      
\begin{tabular}{c c c c c }    
\hline\hline                
 E$_{Line}$ &  $\sigma$ & EW & Flux &  $\chi^2$/dof  \\    
   (keV) & (eV) &  (eV)& ($\times$10$^{-6}$ ph cm$^{-2}$s$^{-1}$) &   \\
\hline 
\\                       
6.39$^{+0.02}_{-0.02}$ & 36$^{+18}_{-24}$  & 121$^{+11}_{-16}$ & 9.62$\pm$1.07&  948/964 \\  

7.02$^{+0.02}_{-0.02}$ &  $<$57 & 52$^{+15}_{-12}$  & 3.61$^{+1.02}_{-0.89}$  & - \\ 

6.70$^{+0.04}_{-0.04}$ &  $<$85 &  18$^{+10}_{-10}$   & 1.61$^{+0.88}_{-0.94}$  & - \\
 
\hline                                   
\end{tabular}
\caption{\label{table:lines}Fe K lines parameters in the 2-12~keV  best-fitting model of the EPIC data.
(Energies are rest-frame, the photon index is $\Gamma$=1.62$^{+0.02}_{-0.02}$. }
\end{table}

When the EPIC spectrum is  fitted in the 0.5-12~keV energy range with a simple power law model, the residuals clearly reveal strong emission in the Fe K complex and emission in excess in the hard X-ray continuum  (Fig.~\ref{fig:epic_ratio}).
The energy range has therefore been restricted  to the 2-12~keV band in order to investigate this region of the spectrum. Fitting the data with a power law of photon index $\sim$~1.6 yields an 
unacceptable fit ($\chi^2$/d.o.f.=1209/971).
A zoom of the ratio of the hard X-ray  data to a power law model is shown in the top panel of  Fig.
\ref{fig:lines}: three emission lines   are clearly seen in all the three detectors
in the $\sim$6-7~keV range.
Three Gaussian emission lines were added, initially  with width free  to vary.
The best-fitting lines parameters are summarised in Table \ref{table:lines}.
 Adding a 6.4~keV line yields an improvement in $\Delta\chi^2$=215, for 3 d.o.f. 
The line width can be constrained for this line to $\sigma$=36$^{+18}_{-24}$~eV. 
  A second line at 7.02$\pm$0.02~keV improves the fit of $\Delta\chi^2$=36, for 2 d.o.f. 
  with an upper limit on the line width  of $\sigma$$<$57~eV.
  The width is thus  kept fixed to 1~eV.
A third line is detected with lower significance at $\sim$6.7~keV with an upper limit on the width of 85~eV  
 ($\Delta\chi^2$=10 for 2 d.o.f.). 
The line width is kept fixed to 1~eV, as for the previous case.
 The confidence contours for these three lines are plotted in the bottom panel of 
 Fig.~\ref{fig:lines}. 

\begin{figure}
   \centering
   \includegraphics[angle=-90,width=8cm]{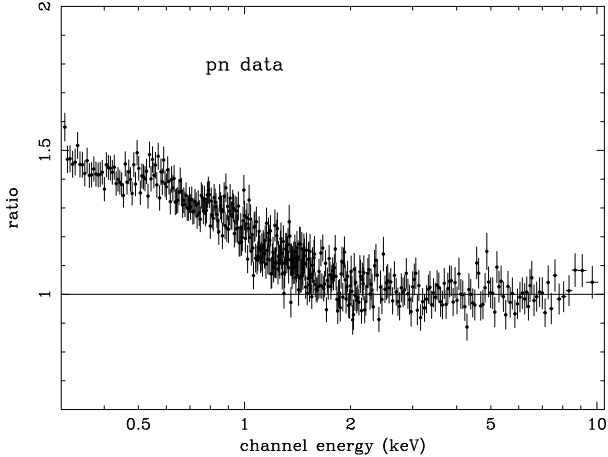}
   \caption{Data to model ratio of the pn spectrum fitted with the hard X-ray best fit in Table 2.  The presence of a soft X-ray component is evident in the figure.}
              \label{soft_excess}%
    \end{figure}

\begin{figure}
   \centering
   \includegraphics[angle=-90,width=8cm]{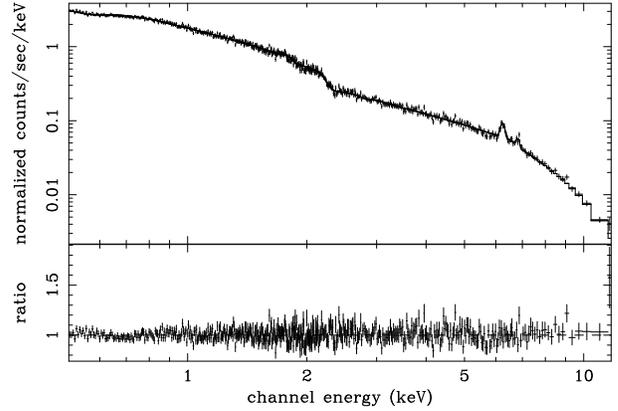}
   \caption{Spectrum and ratio of the pn data fitted with the best-fitting  model (power law, three Fe K emission lines, reflection component and blackbody). }
              \label{Fig:bestfit}%
    \end{figure}

In the following, we analyse the spectrum over the entire
 energy range using the pn data only, since adding the MOS  
 does not yield a relevant improvement for the analysis of the continuum.
 Nonetheless the MOS  have been checked and they are confirmed to  be  consistent with the  pn results.
 The power law+3 gaussian lines fit extended to the full energy range is not satisfactory ($\chi^2$/d.o.f.=1173/750):
 the presence of a soft X-ray component is clear when the data below 2~keV are taken into account (Fig. \ref{soft_excess}).
 The data were checked for the presence of  Compton reflection  since the
 narrow neutral Fe K line is likely to be originated via reflection onto optically thick matter and  the residuals still present above $\sim$5-6~keV indicate that this may be the case.
A {\small PEXRAV} component (Magdziarz \& Zdziarski, 1995) with an inclination angle fixed to 30$^{\circ}$  was thus added to the model. The reflection parameter  is defined as R=$\Omega/2\pi$ with  R$\sim$1 if the reprocessing 
material covers 2$\pi$ of the source.
The addition of the reflection component improves the  $\chi^2$ considerably ($\chi^2$/d.o.f.=811/749), but the fit yields  an extremely high reflection fraction, R$\sim$3.6, and unsatisfactory  residuals in the soft X-ray band.
 We have then tested for the presence of  two additional components i.e. 
 absorption and blackbody emission.
  No absorption seems to be  required 
 by the data and in fact neither cold (model {\small WABS}),  nor warm absorber ({\small ABSORI}) models   improve significantly the fit.
 Adding a blackbody component instead yields  $\chi^2$/d.o.f.=762/747, 
 with best-fitting parameters $\Gamma$=1.73$^{+0.03}_{-0.03}$,  R=1.3$^{+0.7}_{-0.6}$
  and  blackbody temperature kT=156$^{+14}_{-12}$~eV. 
 We note that the blackbody parameters are in good agreement with the results from Gallo et al. 2006. 
 The spectrum fitted with this model and the resulting  residuals  are plotted  in Fig.\ref{Fig:bestfit}.
 The description of the soft excess with a blackbody model is somewhat unsatisfactory.
  The issue of finding a physical explanation for  the soft excess represent a well-known problem in AGNs studies (e.g. Gierlinski \& Done, 2004). It has been shown in fact that the majority of Active Nuclei are characterised by the same temperature (kT$\sim$0.1 keV, see Piconcelli et al. 2005, Crummy et al. 2006)  when their soft X-ray spectra are fitted with a blackbody model. This is  in contradiction with the prediction  that  the blackbody spectrum emitted by an accretion disc associated to a  super-massive black hole  emits at   a much lower    temperature (i.e. $\sim$tens of eV). Alternative models have been tried in the following.
A broken power law with break energy around 1.9-2~keV and with $\Gamma$$_{soft}$=1.89$\pm$0.01 and $\Gamma$$_{hard}$=1.62$\pm$0.01, yields a slightly better fit ($\chi^2$/d.o.f.=758/746). This implies  that the  shape of  the soft X-ray spectrum is not compatible only with a blackbody spectrum.
Recently, the {\small REFLION} model (Ross \& Fabian, 2005) has been successfully employed to fit the soft excess in AGN providing that relativistic blurring is applied  to the spectrum (Crummy et al. 2006). 
  The presence of the ionised Fe lines may suggest an origin in an ionised accretion disc, although  there is no evidence of relativistic effects in Mrk~590, mainly because the Fe  lines profiles are found to be narrow.
 We have tested {\small REFLION} with a power law and a narrow Gaussian line at 6.4~keV (which would not be calculated by  the photoionised disc code).
 This model does not improve the fit statistic ($\chi^2$/d.o.f.=793/750)
 and neither  it does  provide a good fit to  the ionised lines.
 We conclude that ionised reflection and a disc origin for the ionised Fe lines can be excluded in this case. 

%

\subsection{{\it XMM-Newton}/RGS  spectra}
\label{subsec:rgs}
 
\begin{figure}
   \centering
\includegraphics[width=8.5cm]{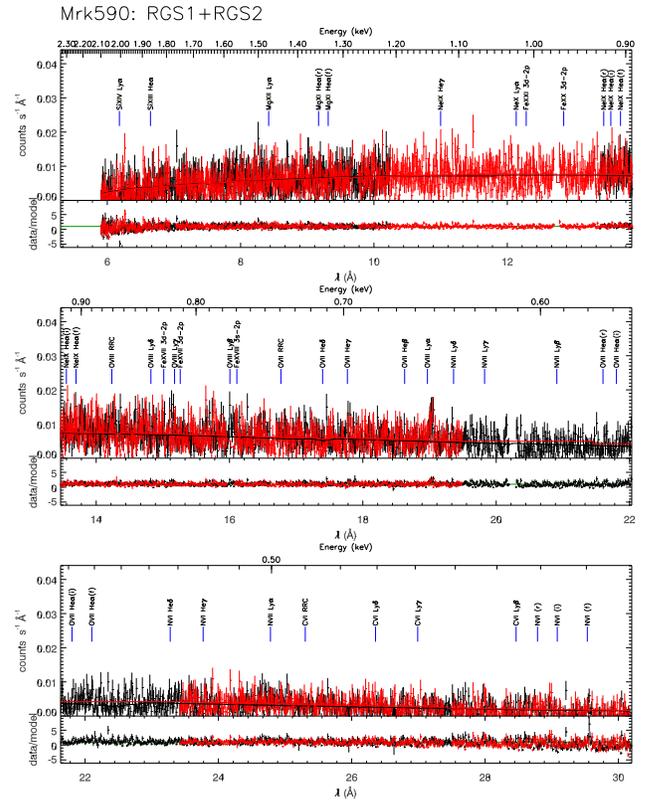}
      \caption{ The plot shows the RGS1 (black) and RGS2 (red) spectrum and the residuals of the data fitted with Galactic absorption and the best fitting  power law with photon index $\Gamma$=1.68$\pm$0.03. The only prominent line  is the OVIII Ly$\alpha$ around 19~$\AA$. This figure has been produced reprocessing the data with SAS 7.0.0 due to the better calibrations.}
         \label{Fig:rgs_stefano}
   \end{figure}

\begin{figure}
   \centering
\includegraphics[angle=-90,width=8cm]{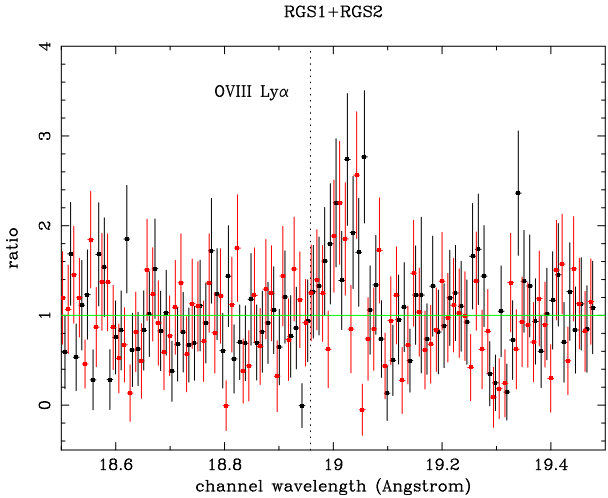}
      \caption{
        Ratio of the RGS data to a power law model in the 18.5-19.5~$\AA$ (corrected for the redshift of the source). The emission line at $\sim$19~$\AA$ is evident in both RGS1 (black points) and RGS2 (red points).  The dashed line marks the OVIII~Ly$\alpha$ rest frame wavelength. }
         \label{Fig:oxy_line}
   \end{figure}

The absence of intrinsic absorption in the soft X-ray spectrum emerged by the analysis 
of the EPIC data, is confirmed by looking at the RGS spectrum (Fig. \ref{Fig:rgs_stefano}).
Indeed no obvious emission or absorption features are visible above the noise, except 
for the OVIII Ly$\alpha$ at $\sim$19~$\AA$.
 The unbinned spectrum  was therefore analysed with {\small XSPEC} in a small wavelength range around this emission line
with the aim  to perform a ``local''  fit.
The C-statistic (Cash 1976) was applied because some energy bins may fall in the limit of few number of counts. Statistical errors are given at 90\% level of confidence ($\Delta$C=2.71).
  The ratio of the data in the $\sim$18-20~$\AA$  to a  model consisting  of a power law and the Galactic absorption, is plotted  in Fig.\ref{Fig:oxy_line}:  a  narrow line is evident in the data.
  When the line is fitted with a  0-width  Gaussian profile, it is detected at $\lambda$=19.02$\pm$0.01~$\AA$ with $\Delta$C=25.5 for 2 d.o.f. The flux is 1.06$^{+0.43}_{-0.48}$$\times$10$^{-5}$ photons~cm$^{-2}$~s$^{-1}$.
The line appears to be emitted with some not negligible shift  ($\sim$$\Delta\lambda$=0.060$\AA$) with respect to the rest wavelength of the OVIII~Ly$\alpha$ i.e. $\lambda$=18.96~$\AA$.    
  
  We have checked the spectrum for the presence of similar shifts in other emission features, but the X-ray continuum is so intense that it may well hamper the detections of other line complexes.
  The only emission structure that may be strong enough to be detected is the OVII triplet at$\sim$22~$\AA$ (see Fig.\ref{Fig:rgs_stefano}). 
Unfortunately,  only the RGS1 data are available in this wavelength range due the 
  electronic chain failure of CCD 4 in RGS2 loss.   
  The upper limits  on the intensity of two lines in the triplet  could be obtained: for the lines at 21.60~$\AA$  and  22.10~$\AA$,  they are respectively 1.16$\times$10$^{-5}$ and 2.05$\times$10$^{-5}$ photons~cm$^{-2}$~s$^{-1}$, but no information on the lines position can be gained. 
  We cannot thus confirm if the shift in the wavelength peak of the OVIII~line is common to other features in the  RGS.  
   
\subsection{The {\it Chandra/HETGs} spectra}
\label{subsec:hetgs}
\begin{figure}[h]
   \centering
\includegraphics[angle=-90,width=7.5cm]{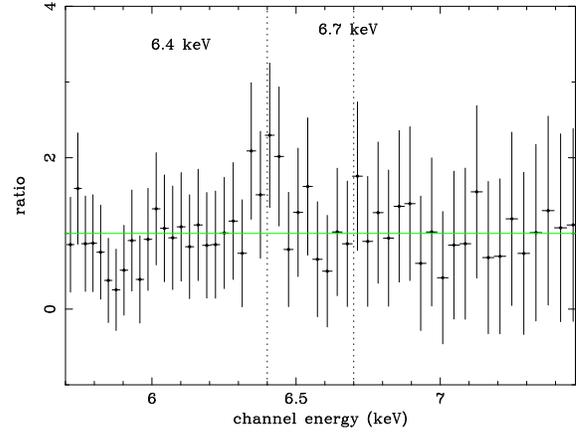}
      \caption{Ratio of the Chandra HEG  data to a power law model (plotted in the source frame).   }
         \label{Fig:chandra_rat}
   \end{figure}

The HEG spectrum was fitted  in the hard X-ray band by a power law with $\Gamma$=1.68$^{+0.10}_{-0.08}$, consistent with the value found in the {\it XMM-Newton} data.
 The ratio of the data to a power law  model in the 5-8~keV is plotted in Fig.~\ref{Fig:chandra_rat}.  Only 190 counts are present in the 6-7~keV range. 
 
 We checked for the presence of the Fe~K lines detected 
in the EPIC spectra.
Adding a narrow line at 6.4~keV improves the fit by  $\Delta$C=22 (for 3 d.o.f.).
The parameters of the line are consistent with the results  in Section~\ref{subsec:epic} from the EPIC data.
For the neutral line E=6.40$^{+0.04}_{-0.03}$~keV, $\sigma$=47$^{+58}_{-24}$~eV and  EW=160$^{+118}_{-78}$~eV.  The line flux is found to be 1.5$^{+1.1}_{-0.7}$$\times$10$^{-5}$~photons~cm$^{-2}$~s$^{-1}$.
Adding another  Gaussian with unresolved width   at $\sim$6.7~keV does not improve 
the fit very significantly ($\Delta$C=3, for 2 d.o.f.) in fact only the upper limit on the line intensity was found 
(7.8$\times$10$^{-5}$~photons~cm$^{-2}$~s$^{-1}$).
 No line is detected at higher energies, but the upper limit on the flux  and corresponding  EW  for a narrow line at 7.0~keV (respectively 2.5$\times$10$^{-6}$~photons~cm$^{-2}$~s$^{-1}$  and 30~eV) are   consistent with the results from {\it XMM-Newton}.

In the lower energy band, the MEG spectrum was searched for the presence of spectral  lines, although the intense continuum would make their detection rather difficult, as already pointed out  for the RGS data (section \ref{subsec:rgs}). 
We concentrate in particular in the range 18-20~$\AA$ with the intent to confirm 
if the  $\sim$~19~$\AA$ line is observed with the same shift as it was in the RGS.
We have performed a local fit  as done for RGS, employing C-statistic  and fitting a zero-width Gaussian line  in a narrow band around the expected emission  wavelength.
The line is actually present in the MEG data and  it is detected with $\Delta$C=16.5 at a wavelength thoroughly consistent with  the rest-frame position:
$\lambda$=18.97$\pm$0.01~$\AA$. The observed flux is 2.10$^{+1.48}_{-1.19}$$\times$10$^{-5}$~photons~cm$^{-2}$~s$^{-1}$.
 No other feature was found in the MEG spectrum.

The difference between the peak wavelength of  the line observed in  the RGS spectrum and in the MEG spectrum is not negligible ($\sim$$\Delta\lambda$=0.060$\AA$).
We investigate the reasons of the discrepancy in the following.

As reported in Table \ref{tab:log} the {\it XMM-Newton} and {\it Chandra} observations
are not exactly simultaneous, but they overlap for a total time of  about 65~ks. 
 Therefore, it was first checked if the discordancy found in the RGS  and  MEG data concerning the detection of the emission line at $\sim$19~$\AA$, could be due to variability intrinsic to the source. The RGS spectrum was extracted from  the portion of time when both satellites were   operative, yielding an exposure of 58~ks.
 After repeating  the analysis  of the line using these data, no change in the line position is found, so it is very unlikely that  the two measurements disagree for physical reasons related to source variability.

An error in the choice of the  source coordinates may introduce a systematic error in the wavelength  assigned to each photon of  the source spectrum; in our case the RGS data were processed inputting the  coordinates corresponding to the optical nucleus of  Mrk~590 so we tend to exclude that the spectrum was shifted due to an erroneous  choice of the source position occurred during the data reduction.
We have checked if this same problem may have occurred in the Chandra data.
In fact if the zero-order image in the Grating event list is piled up, it may lead 
to an incorrect wavelength scale \footnote{See http://cxc.harvard.edu/ciao3.3/threads/tg\_piled\_zero/}. 
Our observation resulted not affected by this issue.
 Once the goodness of the  line position in Chandra  is assured, it seems reasonable  not to question it,  simply because the line is detected at the rest wavelength, i.e. where it is expected. 
  
 The systematic errors in the wavelength scale in  the RGS
are of the order of 20 m$\AA$.
The difference between the laboratory and the  measured wavelengths in a set of certain emission lines is generally around this value, but it may be larger for some 
individual measurements  \footnote{http://xmm.vilspa.esa.es/external/\\xmm\_user\_support/documentation/uhb/node54.html}.
We conclude that the shift in the RGS is probably due to instrumental effects because of  an atypical error in the wavelength  occurred in this observation.

\subsection{{\it XMM-Newton}: Optical Monitor data}
The fluxes from the Optical Monitor were obtained in five bands: B, U, UVW1, UVM2 and UVW2 ( see Table~\ref{tab:om}).
These values  were  estimated by converting  the mean count rates 
in each OM filter, according to the guidelines provided in the SAS documentation.
They have been corrected for extinction assuming the extinction law by Cardelli et al. (1989).
The images in the B and U bands could not be used in order to estimate the contribution of the galaxy in these filters because of instrumental artifacts. 
Nonetheless, it was possible to ascertain the presence of extended emission in these two images. 
 We have fitted  a  2-dimensional Gaussian to the brightest (point-like) sources in the  field of view and to Mrk~590. This is in general a good model to fit emission from a point-like source, but it is not appropriate to model  extended emission. From the comparison of the fits, it was possible to say that  in the  B and U filters  Mrk~590 was  not well fitted  as the other point-like sources. This is due to the fact that the source must be more extended and therefore we can say that  the contribution of the host galaxy is not negligible in B and U bands.
 In the UVW1 filter the image shows that the emission is concentrated in the nucleus
 rather than in the galaxy   and  in the other filters  the contribution from the active nucleus is clearly  predominant. 
 
\begin{figure}[h]
   \centering
\includegraphics[angle=-90,width=7cm]{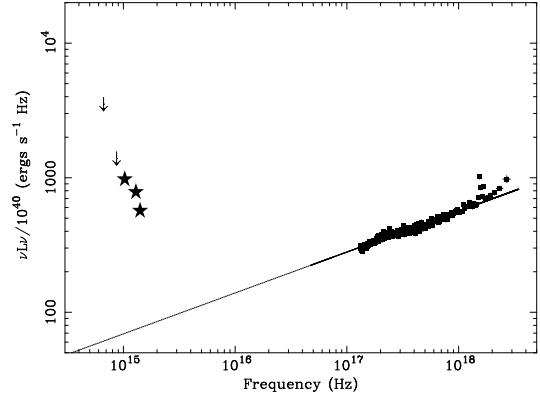}
      \caption{Spectral energy distribution including the X-ray and UV-optical data, with the extrapolation of the best fit X-ray power law.
      The luminosity in the {\it y} axis is in units of 10$^{40}$ ergs/s. The points marked as upper limits are the U and B bands points,  where the contribution of the host galaxy has not been subtracted. The points marked as stars are the other UV fluxes obtained by the OM.}
         \label{Fig:sed}
   \end{figure}

The spectral energy distribution was constructed by taking the OM fluxes and  the X-ray data points in the 0.5-12~keV (Fig.\ref{Fig:sed}).  
 The points  in the U and B bands are marked as upper limits in order to highlight that the  contribution of the host galaxy, although estimated qualitatevely,  is not negligible.
 The fluxes measured in the optical/UV  are located above the extrapolation of the X-ray power law,  indicating the presence of a UV bump. 
 It is  nonetheless quite difficult to relate the shape of this component to the soft X-ray spectrum due to the Galactic absorption between the two bands.
 To clarify the relation between the emission in the  optical/UV and X-ray band, 
 the index $\alpha_{ox}$ defined as the slope of a power law extending between 2500~$\AA$
and 2~keV rest frame can be used as a good indicator of the ``X-ray loudness" of an AGN.
The definition $\alpha_{ox}$ =-0.3838log[F$\nu$(2keV)/F$\nu$(2500$\AA$)]
reported  by Strateva and collaborators (2005) was adapted with the flux at 2310~$\AA$ to replace the UV flux
in the ratio. This change would not affect the results because the fluxes at  2310 and 2910~$\AA$ differ by few\% (see Table~\ref{tab:om}).
The $\alpha_{ox}$ calculated in this way is found to be -0.88, slightly larger than 
the value found for the objects studied by  Strateva et al. 2005, but still typical 
for Seyfert 1 galaxies. 

 \begin{table}       
\centering   
\begin{tabular}{c c c }
\hline\hline                
Filter & $\lambda$  & Flux \\     
-  &  ($\AA$)  &  ($\times$10$^{-15}$ erg cm$^{-2}$ s$^{-1}$$\AA^{-1}$)    \\
\hline 
B & 4500 & 5.76$\pm$0.01 \\
U  & 3440 & 2.97$\pm$0.01 \\
UVW1 & 2910  & 2.46$\pm$0.02 \\
UVM2  & 2310  & 2.49$\pm$0.07 \\
UVW2  & 2120  & 1.97$\pm$0.10 \\                    
 \hline\hline                                   
\end{tabular}
\caption{\label{table:om_fluxes}Fluxes obtained from  the Optical Monitor data. These values are corrected for the Galactic extinction.}
\label{tab:om}
\end{table}

\section{Discussion}

\subsection{The hard X-ray spectrum and the Fe K lines}
The 6.4~keV  Fe~K line detected by both observatories in Mrk~590
appears too narrow to be produced in the inner accretion disc,
otherwise  its profile would be distorted and broadened by the  black hole gravity.
 Consequently,   the line producing material 
 has  to be situated  in outer regions far away from the nucleus.
 One viable interpretation is  that the line is originated   via Compton reflection onto the torus-shaped material distributed at large distance from the central source  according to the AGN Unification Models (Ghisellini, Haardt \& Matt 1994). 
 The reflection fraction determined in our best-fitting model and the EW of the 6.4~keV line are  quite in good agreement with the predictions of George \& Fabian (1991) for 
reflection on optically thick matter.
The results from  the spectral fit in section \ref{subsec:epic} therefore  concur to suggest that the hard X-ray spectrum  arises from reprocessing on the torus.

 Another possible explanation for the Fe~K  line could be emission in  the Broad Lines emitting gas.
The extremely good quality of the {\it XMM-Newton} spectrum  combined with the {\it Chandra} grating allowed to measure the width of the Fe~K line in this source. 
The two measurements are in good agreement ($\sigma$=36$^{+18}_{-24}$~eV in EPIC and $\sigma$=47$^{+58}_{-24}$~eV in the HEG data),  corresponding to a FWHM  of 4000$^{+2000}_{-2700}$~km~s$^{-1}$  and 5250$^{+6500}_{-2700}$~km~s$^{-1}$ respectively. 
The FWHM of the H$\alpha$ line in Mrk~590 is reported  to be  $\sim$~2400~km~s$^{-1}$ (Stirpe 1990), implying that this source is actually much closer to the Narrow Line Sy~1 class (Boller, Brandt \& Fink 1996), at least with respect to  the optical lines  properties.
Unfortunately, this means  that  regardless of  resolving  the  Fe K$\alpha$ line width as done herein, it is somewhat difficult  to ascertain  if the line comes from the BLR because 
the width of the BLR lines in Mrk~590 is  smaller than in the typical Sy1 case.      
The torus hypothesis,  corroborated by the presence of the neutral  reflection component in the spectrum,  remains therefore the favoured scenario for the production of the Fe~K line.
\begin{figure*}
\begin{center}
\centerline{
\resizebox{8cm}{!}{\psfig{figure=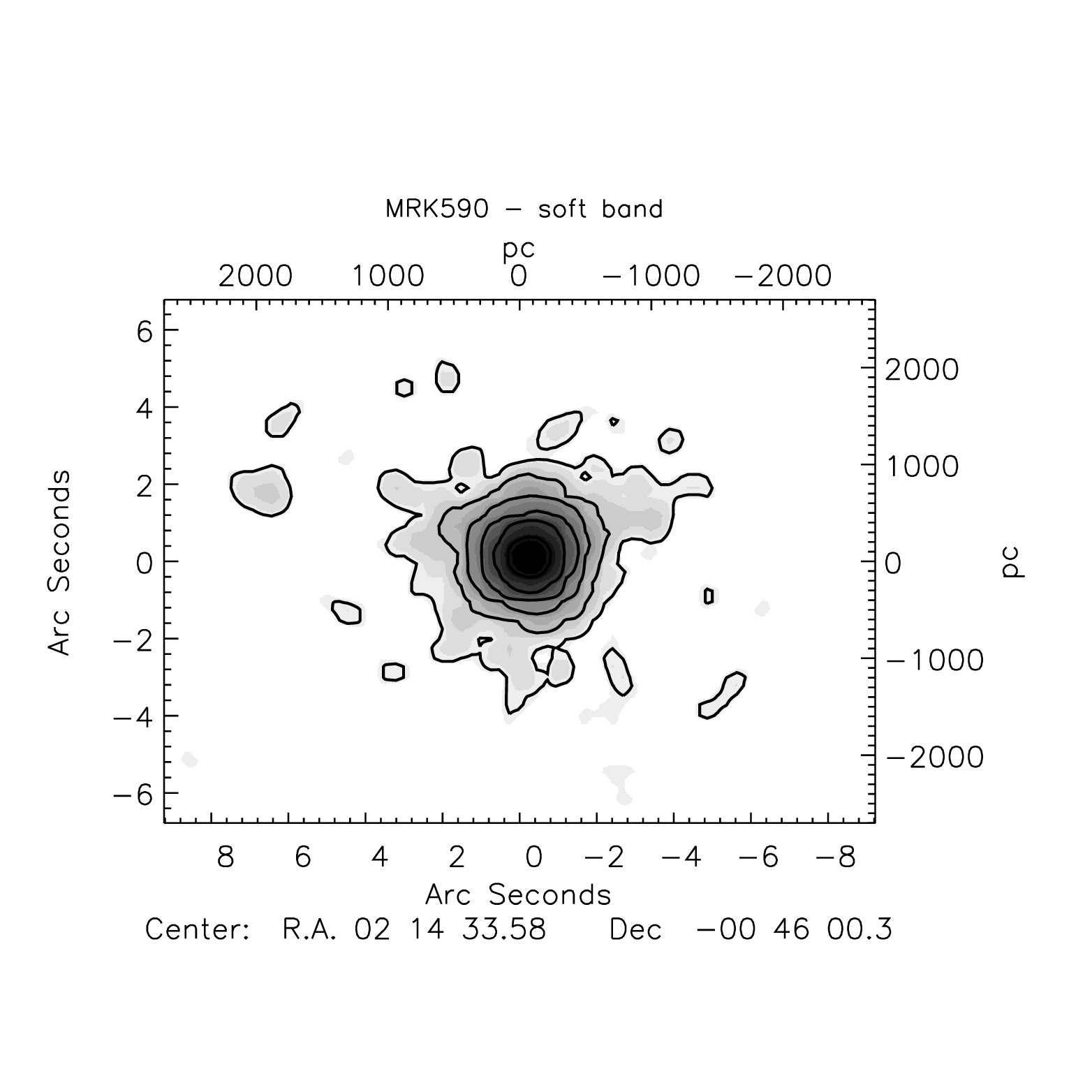}}
\hspace{0.15 cm}
\resizebox{8cm}{!}{\psfig{figure=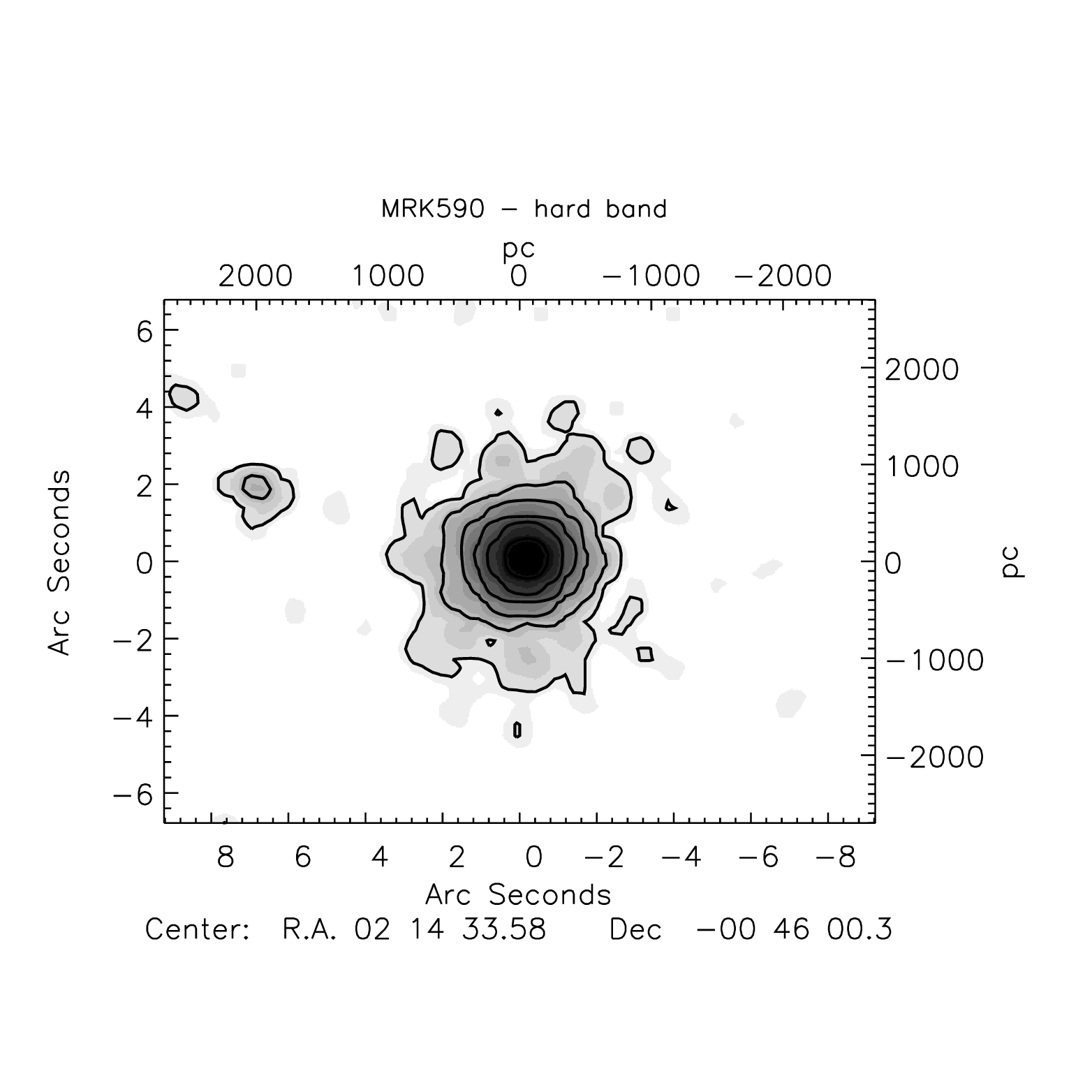}}
}
\caption{{\it Chandra} images of Mrk~590 in  the 0.3-2~keV band ({\it Left})   and in the 
2-10~keV band ({\it Right}) with contour levels of the emission superimposed. 
  The source  visible to the west of the nucleus of the AGN (Source A in this paper) is discussed in  Section~\ref{sec:serendipity}.  }   
\label{Fig:im_sourceA}
\end{center} 
\end{figure*}

 The two lines emitted at $\sim$6.7 and $\sim$7~keV  are fully consistent  with emission from Fe XXV (6.7~keV) and XXVI (6.96~keV), therefore from Iron in a very ionised state.
 A contribution of the K$\beta$ from neutral Fe to the Fe~XXVI line cannot be ruled out by these data. The centroid energy of the line is in facts slightly 
higher ($\sim$7.02~keV) than what would be expected for Fe XXVI only. 
The 7~keV  line is not detected in Chandra and the addition of a  
 K$\beta$ component in the EPIC best fit (Table~\ref{table:lines})  was not significant.
 Nevertheless, it is possible to obtain a qualitative estimate of 
  the contribution of the K$\beta$  to the XXVI~Fe line taking into account 
  the flux ratio  Fe~K$\beta$/Fe~K$\alpha$, which has been calculated to be around 0.15 for neutral Iron (see Molendi et al. 2003).
  Referring to this value,  we have included a fourth Gaussian line to the  best fit reported 
 in Table~\ref{table:lines} with energy fixed to 7.06~keV and intensity fixed to 1.34$\times$10$^{-6}$~photons~cm$^{-2}$~s$^{-1}$, i.e.   the  energy and the estimated  Fe~K$\beta$ flux.
 In this way, the Fe~XXVI line  flux decreases to 2.44$^{+0.94}_{-0.95}$$\times$10$^{-6}$~photons~cm$^{-2}$~s$^{-1}$, corresponding to $\sim$35~eV.
   Ionised Iron lines with EW of tens of eV against the total continuum can  be produced by fluorescence and scattering in a photoionised medium out of the line of sight, not only in  Seyfert~1 objects as  NGC~7213 (Bianchi et al. 2003a) and NGC~5506 (Bianchi et al.2003b),
but also in many Seyfert~2 cases (Bianchi et al. 2005 and references therein).
The {\it Chandra} images  in Fig.~\ref{Fig:im_sourceA} tentatively show that the X-ray  
 emission may be extended on a kpc scale.  The ionised lines may be  
 be originated  in this extended gas, which is not otherwise observable in the spectrum 
 due to the intensity of the  primary nuclear continuum.

The results obtained from the analysis of the hard X-ray spectrum do not 
 provide any particular evidence for   accretion disc signature, in agreement with many  
 observations of type 1 AGNs (Bianchi et al. 2004, Matt et al. 2006).  

 \subsection{The soft X-ray spectrum}
As mentioned in section \ref{subsec:epic}, the spectrum of Mkn~590 is not absorbed, 
making  this source  falling  into  the fraction  of Sy1 Galaxies ($\sim$50\%) that do not show intrinsic absorption   (Reynolds, 1997, George et al. 1998). 
The only relevant feature observed in the  high resolution spectra
 is the O~VIII~Ly$\alpha$ emission line. 
 The line arises with most probability by photoionisation processes 
 in warm gas.
Netzer et al. (2005) provided photoionisation calculations of line ratios  for the case of the starburst galaxy NGC~6240. Among these, the intensities of the Fe XXV and XXVI lines relative to the OVIII Ly$\alpha$ were 
estimated (the line fluxes were quoted in units of ergs~cm$^{-2}$~s$^{-1}$). Using the same units, the line ratios  FeXXV/OVIII and FeXXVI/OVIII in our data are estimated  around 1.6 and 1.3 (this last value is approximated because an exact estimate of the line flux is affected by the presence of the FeK$\beta$). A qualitative comparison with Netzer's results  shows that  it is possible to have photoionised gas producing OVIII Ly$\alpha$ and  strong Fe XXV and XXVI lines with these ratios, providing that the gas is characterised by a rather high value of the ionisation parameter, i.e. U$>$ 2 where U is defined as the photon to hydrogen density ratio in the range 0.54-10~keV.  Of course, this cannot exclude {\it a priori} the  more realistic picture of a multi-phase medium with various ionisation parameters.
 Likewise, the  apparent lack of features from other transitions in the soft X-ray spectrum may indicate that the material is highly 
photoionised  but this is a tentative conclusion since we are unsure about the detection  of other lines in the RGS spectrum (e.g. the OVII triplet, section~\ref{subsec:rgs}). 

 Most of the  Seyfert~1 Galaxies  with narrow emission  in RGS spectra are also characterised by the presence of warm absorbers: for example the OVIII~Ly$\alpha$ emission line has been observed in   NGC~3783 (Behar et al. 2003), NGC~7469 (Blustin et al. 2003) and Mrk~509 (Smith et al. 2006). Particularly, in the case of NGC3783 it was suggested to relate the X-ray absorbing gas with the OVIII line emitting plasma.  Mrk~590 represents a  peculiar case because of  the absence of warm absorption and of the presence of X-ray emission  lines from highly ionised elements at once. We can try to explain the unusual spectrum of this galaxy with simple considerations. The soft X-ray spectrum  appears to be  made by an intense nuclear component  with a mild contribution of a photoionised medium probably in different ionisation  phases  revealed  by the observation of  highly ionised O and Fe  lines. The characteristics of this  medium resemble those found in Seyfert~2 objects, where the soft X-ray spectrum is due to photoionisation of circumnuclear gas (Guainazzi \& Bianchi 2006). Let us assume an ionised cone geometry as for NGC~1068 (see Kinkhabwala et al. 2002).  The lack of warm absorption in Mrk~590 implies that our line of sight does not intersect warm gas, or either that the gas in the line of sight is fully ionised and therefore it does not imprint absorption features.  Thanks to this ``clean'' line of sight,  we can see in this Seyfert~1 galaxy the scattering gas  responsible for line emission in Seyfert~2.
The observed EW of the emission lines in Mrk~590 are weaker than in Seyfert~2 because we are observing directly the active nucleus and they are diluted by the X-ray continuum. 
This scenario provides  yet another confirmation of  the Sy1/Sy2 dichotomy to be due to orientation effects.

 \section{Analysis of the serendipitous  X-ray sources in the field of view}
 \label{sec:serendipity}
 \begin{table}      
\centering                   
\begin{tabular}{c c c c}    
\hline\hline                
Source  & RA  & DEC  & Flux (0.3-10~keV) \\     
-  &   - &  -  & ($\times$10$^{-14}$ erg cm$^{-2}$ s$^{-1}$)    \\
\hline 
\\
A     & 02h 14m 34.0s   &  -00h 45m 58s     &  2.88$^{+2.5}_{-1.8}$ (ACIS) \\

 B  & 02h 14m 34.7s  &  -00h 42m 43s         & 26.7$^{+0.9}_{-0.8}$  (MOS-2) \\
 
C    & 02h 14m 36.4s  &   -00h 42m 58s         & 2.68$^{+0.42}_{-0.38}$ (MOS-2) \\
\\
                 
 \hline\hline                                   
\end{tabular}
\caption{List of the serendipitous sources detected in the field of view of {\it Chandra} (source A, B, C) and {\it XMM-Newton} (source B and C). Source B is already know as 1WGA J0214.5-0042.}
\label{tab:sources}
\end{table}

Besides  the main target of these two observations,  
we have also studied  three X-ray sources detected in the field of view of  the {\it XMM-Newton} and {\it Chandra}  images.
For clarity, they will be referred to as Source A, B, C.
Table~\ref{tab:sources} lists their positions and fluxes.

\begin{figure}
   \centering
   \includegraphics[width=8.5cm]{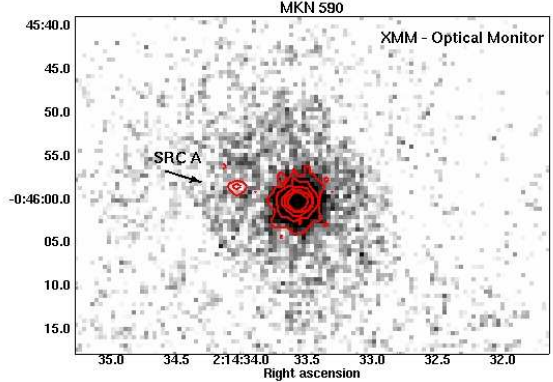}
      \caption{ XMM-Optical Monitor image in the UVW1 filter of Mrk~590 with Chandra contours overlapped. The image has been co-aligned with the one obtained by Chandra in order to take into account  the different  astrometry error of the two instruments. As shown, no counterpart is evident for Source A.}
         \label{Fig:OM_590}
   \end{figure}

\begin{figure}
   \centering
   \includegraphics[width=8.5cm]{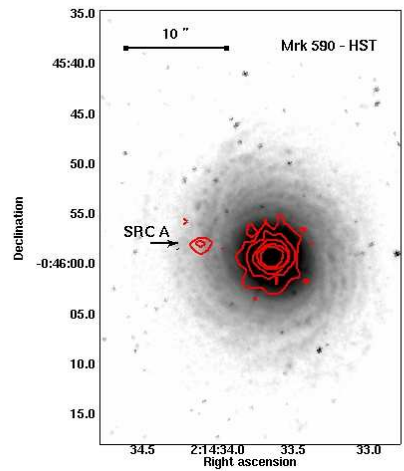}
      \caption{ HST image of Mrk590 obtained in 1995. The overlapped contours mark the X-ray emission from Chandra. At the position of  Source A no point source appears to be present.  The X-ray source seems  to be located within the extension of the galaxy and this may be favouring the hypothesis that Source A could be  identified as a ULX in the galaxy. The HST and Chandra images have been co-aligned at the position of the X-ray nucleus.}
      \label{Fig:hst_590}
   \end{figure}

Source A   is visible  only in the {\it Chandra} image at $\sim$~7~arcsec northwest of the nucleus, thanks to the high spatial resolution of the satellite.
Figure~\ref{Fig:im_sourceA} shows the contour levels of the emission in the soft X-ray band and in the hard X-ray band, where Source A seems to be stronger.
The ACIS spectrum was extracted yielding 46~counts in the 0.3-10~keV.
When it is fitted with a power law and  Galactic absorption,  
the photon index is found to be $\Gamma$=1.13$\pm$0.5.
The observed  0.3-10~keV flux is  $\sim$2.88$\times$10$^{-14}$~ergs~cm$^{-2}~s^{-1}$, corresponding to a luminosity of about 4.3$\times$10$^{40}$~ergs~s$^{-1}$ if the source is at the same redshift of Mrk590. 
We have searched in the OM images if any possible optical/UV counterpart is visible.
There is no clear excess at the source position in none of them.
  From the position of the {\it Chandra} contours overlapped to the  UVW1 image 
(Figure \ref{Fig:OM_590}),  we can deduce that Source A  seems located in the host galaxy.
We also have searched for  archival images of Mrk~590 obtained  by the Hubble Space Telescope: Figure~\ref{Fig:hst_590} shows Mrk~590 as observed by the Wide Field Planetary Camera 2 (WFPC2)
with the filter F606W applied. The observation was performed on June 24th~1995 for 500~s.
Figure~\ref{Fig:hst_590} shows  the {\it Chandra} contours  overlapped to the optical image and although no optical counterpart is clearly visible,  the location of   Source A indicates that it could be a Ultra Luminous X-ray source  in the spiral arms of Mrk~590.
If the identification was confirmed,  Source A represents a fairly  typical ULX, although 
it must be noted that only  few ULX of those catalogued and studied so far, exceed  luminosities of  3-4$\times$10$^{40}$~ergs~s$^{-1}$ (Liu \& Mirabel, 2005, Miniutti et al. 2006). 

Another possibility for the origin of Source A  is that it could be a background AGN. Assuming that the present field  is characterised by the same number of  expected X-ray sources within  a given area as observed by the deep {\it XMM-Newton} observation of the Lockman Hole, we can estimate the probability of Source A to be an AGN from the  Lockman Hole {\it Log N-Log S}  (Hasinger et al. 2001). 
 According to the {\it Log N-Log S}, 
 at  fluxes equal or higher than  5$\times$10$^{-15}$~ergs~cm$^{-2}~s^{-1}$ in the 0.5-2~keV band, (corresponding to Source A flux),  200 X-ray sources per square degree are expected.
 Considering a  circular  area of radius equal to the distance of Source~A from Mrk~590 
 ($\sim$7.3~arcsec), the number of expected X-ray sources is 0.002 in an area of 1.256$\times$10$^{-5}$deg$^2$. 
 The use of the  hard X-ray flux yields a similar estimate,  i.e. 0.001 for the same area.
 These numbers  indicate that the probability  for Source A to be a background AGN is  small but not null and therefore the origin of this source remains open.

Source B  and Source C  are visible in the EPIC-pn image of the field of view but 
the exact  positions are taken from {\it Chandra} where the sources are clearly separated.
We have found two possible  optical counterparts in the Digitized Sky Survey, where 
the two sources appear extremely faint.

\begin{figure}
   \centering
   \includegraphics[width=6cm,angle=-90]{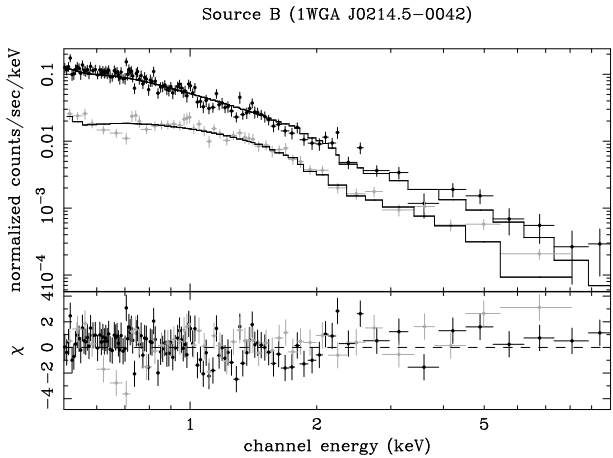}
      \caption{  Pn (black points) and MOS-2 (grey points) data of Source B (1WGA J0214.5-0042). 
The spectra in the top panel are fitted with a power law of $\Gamma$$\sim$2.4 and the bottom panel shows 
 the residuals of this fit.
      }
         \label{fig:pn_image}
   \end{figure}

Source B was already known as 1WGA J0214.5-0042, detected by ROSAT
in the soft X-rays (White, Giommi \& Angelini, 2000). 
We have searched if the source is present in other catalogues or if other observations in different wavelengths are available, but only ROSAT has detected  this source.
Unfortunately, Source B falls  just out  of the field of view of the OM because of the orientation of the telescope, and out of the field of view of the MOS-1 because of the choice of the small window mode. Therefore, only the X-ray data from the pn and MOS-2 can reveal something on the nature of this source. 
The {\it XMM-Newton} spectrum is shown in  Figure~\ref{fig:pn_image}.
We have fitted pn and MOS-2 together with a power law and Galactic 
absorption towards the direction of Source B.
The resulting photon index in the range 0.5-10~keV is quite steep, $\Gamma$=2.39$\pm$0.05 with $\chi^2$/d.o.f.=415/356.
The fluxes are 1.71$\pm$0.05$\times$10$^{-13}$~ergs~cm$^{-2}~s^{-1}$
in the 0.3-2~keV and 9.55$^{+0.32}_{0.51}$$\times$10$^{-14}$~ergs~cm$^{-2}~s^{-1}$.
If the data are instead fitted with a broken power law with break energy
E=2~keV, the slopes are $\Gamma_{0.3-2}$=2.48$\pm$0.05 and $\Gamma_{2-10}$=1.89$\pm$0.15.

Source C is not present in any catalogue.
The spectra extracted from the pn and MOS-2 and fitted with a power law and Galactic absorption yield $\Gamma$=2.02$\pm$0.23, in the 0.5-10~keV range.
The observed soft X-ray flux is  about 1.23$\times$10$^{-14}$~ergs~cm$^{-2}~s^{-1}$ 
and the hard X-ray flux is 1.47$\times$10$^{-14}$~ergs~cm$^{-2}~s^{-1}$.

The lack of optical counterparts and other multiwavelength information on Source B and C make very difficult to understand the nature of these sources.
It is clear from the spectral fitting and the different photon index 
that the two sources are distinct and not related to each other. Moreover as said above, they appear clearly separated in the {\it Chandra} image. 
Taking into account that in both sources the spectra are well fitted by steep power law,    the most likely interpretation for both of them is in term of a background  AGN.

\section{Summary}
 The analysis of the simultaneous {\it XMM-Newton/Chandra} observation of the Seyfert 1 galaxy Mrk~590 has been presented.
In the following the main results of this work are summarised:
\begin{itemize}
\item The analysis of the X-ray data reveals a fairly typical Sy1 spectrum. The broad band continuum is  well described by a 
power law with $\Gamma$$\sim$1.7, a cold reflection component originating in distant matter  and a blackbody 
 with temperature kT$\sim$0.1~keV. 
\item The Iron K complex composed by 3 emission lines at $\sim$ 6.4, 6.7 and 7~keV  has been clearly detected in the EPIC data. The width of the 6.4 line is resolved in both {\it Chandra} and {\it XMM-Newton} spectra with velocities in the range of  4-5000~km~s$^{-1}$.  
The line is interpreted as being originated via reflection onto the torus-shaped material surrounding 
the active nucleus, although an origin in the BLR cannot be excluded.
\item  The two Fe lines at higher energies are consistent with being emitted by photoionisation 
 of FeXXV and FeXXVI. 
The analysis of the high resolution data  highlighted the presence of an OVIII~Ly$\alpha$ emission line
 consistent to be emitted in the same  photoionised gas.
The observation of highly ionised emission lines with no absorption from warm gas is interpreted as an orientation effect: the nucleus is surrounded by warm scattering clouds but it is  seen along a particular  line of sight with either fully ionised material or no intervening material at all.  
\item The OM measurements of the UV fluxes are found to be above the extrapolation of the X-ray spectrum, 
suggesting the presence of a UV bump.
\item Three serendipitous sources are present in the field of view of {\it Chandra} (Source A, B, C) and {\it XMM-newton}/EPIC-pn (only source B and C). All of them are characterised by X-ray power laws and fluxes typical of  AGN.
With regard to  Source A, the location and the estimated luminosity of the source (assuming it belongs to Mrk~590), may induce to interpret it as  a  ULX source, although no counterparts are observed in  optical and ultraviolet data.   
   
\end{itemize}

\begin{acknowledgements}
This paper is based on observations obtained with {\it XMM-Newton}, an ESA science mission with instruments and contributions directly funded by ESA Member States and NASA.
We would like to thank the Chandra X-ray Center and the anonymous referee for helpful comments on this publication.
\end{acknowledgements}

\end{document}